\documentclass[a4paper,UKenglish,cleveref,autoref,thm-restate]{lipics-v2021}

\hideLIPIcs  

\usepackage{pdfpages}

\title{Level~Planarity~Is~More~Difficult~Than~We~Thought} 

\author{Simon D.~Fink}{Algorithms and Complexity Group, Technische Universität Wien, Austria}{sfink@ac.tuwien.ac.at}{https://orcid.org/0000-0002-2754-1195}{}

\author{Matthias Pfretzschner}{Faculty of Computer Science and Mathematics, University of Passau, Germany}{pfretzschner@fim.uni-passau.de}{https://orcid.org/0000-0002-5378-1694}{}

\author{Ignaz Rutter}{Faculty of Computer Science and Mathematics, University of Passau, Germany}{rutter@fim.uni-passau.de}{https://orcid.org/0000-0002-3794-4406}{}

\author{Peter Stumpf}{Charles University, Prague, Czech Republic}{stumpf@kam.mff.cuni.cz}{https://orcid.org/0000-0003-0531-9769}{}

\authorrunning{S. D. Fink, M. Pfretzschner, I. Rutter, and P. Stumpf} 

\Copyright{Simon D.~Fink, Matthias Pfretzschner, Ignaz Rutter, and Peter Stumpf} 



\category{} 

\nolinenumbers 

\EventEditors{John Q. Open and Joan R. Access}
\EventNoEds{2}
\EventLongTitle{42nd Conference on Very Important Topics (CVIT 2016)}
\EventShortTitle{CVIT 2016}
\EventAcronym{CVIT}
\EventYear{2016}
\EventDate{December 24--27, 2016}
\EventLocation{Little Whinging, United Kingdom}
\EventLogo{}
\SeriesVolume{42}
\ArticleNo{23}

\graphicspath{{./figures/}}

\begin{document}

\maketitle

\begin{abstract}
We consider three simple quadratic time algorithms for the problem \textsc{Level Planarity} and give a level-planar instance that they either falsely report as negative or for which they output a drawing that is not level planar.
\end{abstract}

\section{Introduction}
Given a graph $G=(V,E)$ and a \emph{level} assignment $\ell:V(G)\to\mathbb{N}$, the problem \textsc{Level Planarity} asks for a crossing-free drawing of $G$ where the vertices have their prescribed level as $y$-coordinate and all edges are $y$-monotone.
When initially considering the problem in 1988, Di Battista and Nardelli~\cite{DiBattista1988} gave a linear-time algorithm for the restricted case where the graph is a hierarchy, i.e., only one vertex has no neighbors on a lower level.
A subsequent attempt to extend this algorithm to the general case~\cite{Heath1995} was shown to be incomplete~\cite{Juenger1997}.
Jünger et al. finally gave the first linear-time algorithm for testing~\cite{Juenger1998} and embedding~\cite{Juenger1999,Juenger2002} level graphs around the turn of the millennium.
Because this algorithm is quite involved, slower but simpler algorithms were developed by Randerath et al.~\cite{Randerath2001}, Healy and Kuusik~\cite{Healy2004}, as well as Harrigan and Healy~\cite{Harrigan2007} in the decade thereafter.
All these algorithms consider the pairwise ordering of vertices on the same level, greedily fixing an order for a (certain) pair and then checking for further orders implied by this.
If the process terminates without finding a contradiction, we obtain a total vertex order for each level and thereby a level planar embedding. 
In the following, we give a level-planar counterexample that each known variant of this algorithm either incorrectly classifies as negative instance or correctly identifies as positive instance but outputs a drawing that is not level planar.
To the best of our knowledge, this leaves no correct \emph{simple} embedding algorithm for level graphs.
In particular, we are not aware of \emph{any} correct implementation for embedding level-planar graphs.

Randerath et al.\ use an explicit 2-SAT formulation for the pairwise orders of vertices on the same level.
Due to known gaps in the proof of Randerath et al., Brückner et al.~\cite{Brueckner2022} showed this characterization via a 2-SAT formula is equivalent to the Hanani-Tutte-style characterization of \textsc{Level Planarity}~\cite{Fulek2013}.
Thereby, our counterexample only breaks the proof of correctness as well as the embedder by Randerath et al., while their 2-SAT formulation still yields a correct \emph{test} for \textsc{Level Planarity} via this indirect proof~\cite{Brueckner2022}.

\begin{figure}[h!]
	\vspace{-.1cm}
	\centering
	\begin{subfigure}[t]{0.3\textwidth}
		\centering
		\includegraphics[page=1, scale=0.9]{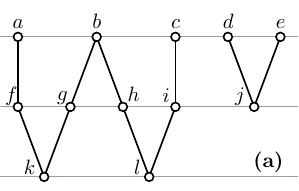}
		\phantomsubcaption
		\label{fig:fig-a}
	\end{subfigure}
	\hfill
	\begin{subfigure}[t]{0.3\textwidth}
		\centering
		\includegraphics[page=2, scale=0.9]{randerath}
		\phantomsubcaption
		\label{fig:fig-b}
	\end{subfigure}
	\hfill
	\begin{subfigure}[t]{0.32\textwidth}
		\centering
		\includegraphics[page=3, scale=0.9]{randerath}
		\phantomsubcaption
		\label{fig:fig-c}
	\end{subfigure}
	\vspace{-.6cm}
	\caption{\textbf{(a)} A level-planar graph $G$. \textbf{(b)} The green, blue, and red 2-SAT equivalence classes can be greedily assigned in this order. Subsequently, transitive closure forces $a < b$ as well as $i < g$, but the planarity constraints force $a < b \leftrightarrow f < h \leftrightarrow k < l \leftrightarrow g < i$ \textbf{(c)}, yielding a contradiction.}
  \label{fig:randerath}
\end{figure}

\section{Randerath et al.}
The algorithm by Randerath et al.~\cite{Randerath2001} works as follows.
First, edges spanning multiple levels are subdivided such that subsequently edges only occur between adjacent levels, resulting in a \emph{proper} level graph.
The planarity of the resulting graph is then tested using a 2-SAT formula.
The formula contains a variable $(a<b)$ for every pair $a,b$ of vertices that appear on the same level, encoding the relative order of these two vertices.
For every pair of edges $uv, xy\in E$ with $\ell(u)=\ell(x)=\ell(v)+1=\ell(y)+1$ with $u\neq x, v\neq y$ it adds the 2-SAT constraint $(u<x) \Leftrightarrow (v<y)$.
Combining this with the constraints for antisymmetry ($(a<b)\Leftrightarrow\lnot(b<a)$) and transitivity ($(a<b)\wedge(b<c)\Rightarrow(a<c)$) necessary for finding total orders yields a 3-SAT formula.
However, Randerath et al.~\cite{Randerath2001} show that omitting the transitivity constraints yields an \emph{equisatisfiable} 2-SAT formula.
To prove this equivalence, they show that the 2-SAT formula can be used to compute a level-planar embedding of the input graph.
They greedily pick and assign equivalence classes of the formula in arbitrary order, but prioritize transitive closures where possible.
\Cref{fig:randerath} shows a counterexample where the algorithm gives a false-negative answer when assigning classes in the shown order.

\begin{figure}[b]
  \centering
  \includegraphics[width=\linewidth]{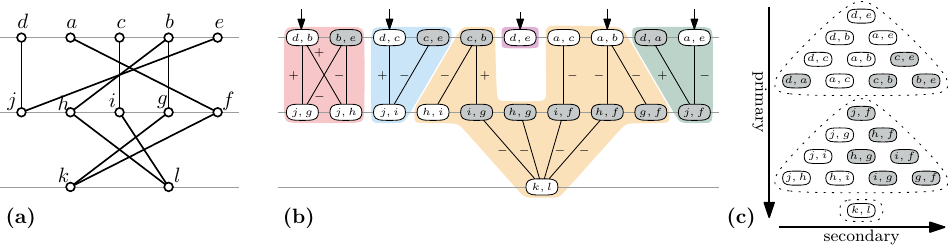}
  \vspace{-.6cm}
  \caption{\textbf{(a)} An initial drawing $\mathcal L$ for \Cref{fig:fig-a}. \textbf{(b)} The corresponding labeled ve-graph. Arrows mark the chose DFS entry points, pairs marked as swapped by Algorithm 1 are shown in gray. \textbf{(c)} The processing order for the vertices of the ve-graph in Algorithm 2.}
  \label{fig:harrigan}
\end{figure}

\section{Healy and Kuusik \& Harrigan and Healy}
The algorithms by Healy and Kuusik~\cite{Healy2004} as well as the one by Harrigan and Healy~\cite{Harrigan2007} uses a similar concept.
Instead of working with equivalence classes of a 2-SAT formula, they work with connected components of the closely related \emph{vertex exchange graph} (ve-graph).
This graph contains one vertex for every ordered pair of vertices that appear on the same level.
Two vertices of the ve-graph are adjacent if they correspond to a pair of independent edges between the same levels.
Starting with an arbitrary drawing $\mathcal L$ of the input graph, the edges of the ve-graph are first labeled with $+$ or $-$, depending on whether the corresponding edges cross in $\mathcal L$.
Subsequently, a DFS is used to test the ve-graph for odd-labeled cycles, which corresponds to a contradiction within a 2-SAT equivalence class.
The two algorithms now differ slightly in how they continue to construct an embedding.
Similar to Randerath et al., Healy and Kuusik~\cite{Healy2004} fix the orders of vertex pairs (i.e., whether all pairs of a connected ve-graph component are swapped or not) in an arbitrary order, also performing the transitive closure if possible.
Thus, the processing order from \Cref{fig:randerath} also breaks this approach. 

\enlargethispage{1em}
The later Harrigan and Healy approach~\cite{Harrigan2007} is slightly more involved.
During the DFS traversal, they already change the relative order of some vertex pairs compared to the initial drawing $\mathcal L$~\cite[Algorithm 1]{Harrigan2007}.
Subsequently, the ve-graph is traversed again in a specific order and, for some vertices of $\mathcal L$, the chosen vertex order is flipped~\cite[Algorithm 2]{Harrigan2007}.
Using choices as in \Cref{fig:harrigan}, this does not yield a planar embedding even for a positive instance.




\bibliography{references}



\section*{Supplementary material}

# Level Planarity Is More Difficult Than We Thought

Simon D. Fink[1] ● Matthias Pfretzschner[2] ● Ignaz Rutter[2] ● Peter Stumpf[3]

[1] TECHNISCHE UNIVERSITÄT WIEN
[2] UNIVERSITÄT PASSAU — Fakultät für Informatik und Mathematik
[3] Charles University

## Level Planarity

**Input:**
Graph $G = (V, E)$
Leveling $\ell : V \to \mathbb{N}$

**Problem:**
Does $G$ admit a planar drawing where
- each vertex $v$ has $y$-coordinate $\ell(v)$
- edges are $y$-monotone

subdivide long edges to make inst. **proper**

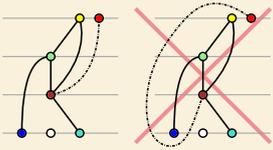
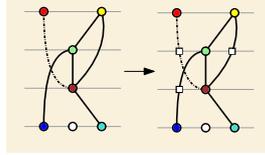

## Algorithms

Jünger et al. ('98, '02): involved $O(n)$ algorithm

↓

subsequent slower but simpler algorithms

| Randerath et al. 2001 [6] | Healy & Kuusik 2004 [3] | Harrigan & Healy 2007 [2] |
|---|---|---|
| $O(|V|^2)$ | $O(|V|^3)$ | $O(|V|^2)$ |

## Counterexample

proper level graph $G$

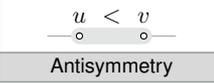

### Randerath. et al [6]:

**Equivalent 3-SAT Formulation:**
- one variable $(a < b)$ for every ordered pair $a, b$ of vertices on the same level
- constraints:

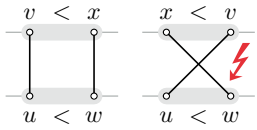

for independent edges $vu, xw$:
$(v < x) \Leftrightarrow (u < w)$

**Planarity**

$(u < v) \Leftrightarrow \neg(v < u)$

**Antisymmetry**

$(a < b) \wedge (b < c) \Rightarrow (a < c)$

**Transitivity**

### Theorem A [1, 6]

2-SAT formula w/o Transitivity satisfiable $\Leftrightarrow$ graph is level planar

**Proof [6]:** greedily pick and assign arbitrary equivalence classes, but perform transitive closure when necessary

Claim: resulting embedding is level planar

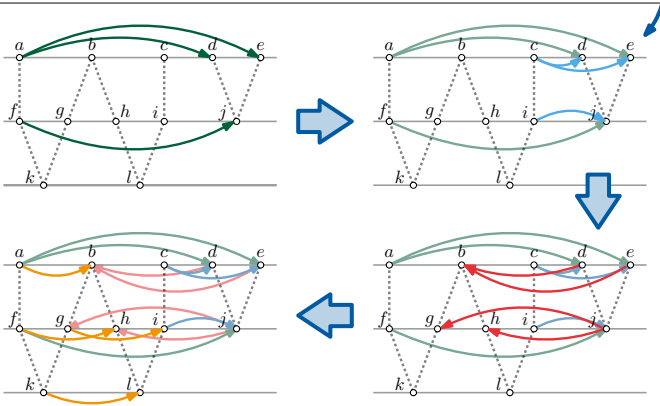

- $a < d < b$ transitively forces $a < b$
- $i < j < g$ transitively forces $i < g$
- planarity forces $a < b \Leftrightarrow f < h \Leftrightarrow k < l \Leftrightarrow g < i$

⚡ contradiction! → false-negative answer

### Harrigan & Healy [2]:
(analogously for Healy & Kuusik [3])

start with arbitrary drawing $\mathcal{L}$:

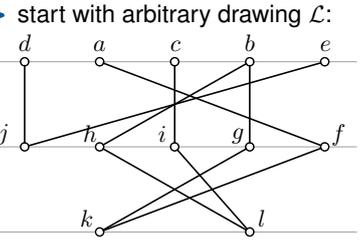

Construct **Vertex-Exchange Graph** $\mathcal{VE}$:
- one vertex for every ordered pair of vertices on the same level
- for independent edges $vu, xw$: connect vertices $v, x$ and $u, w$; if they cross in $\mathcal{L}$ label edge $+$, else label it $-$

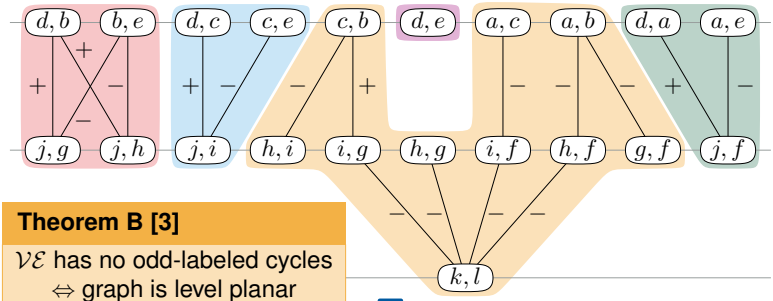

### Theorem B [3]

$\mathcal{VE}$ has no odd-labeled cycles $\Leftrightarrow$ graph is level planar

**Algorithm 1 [2]**
1. **for** each connected component in $\mathcal{VE}$ **do**
2.   fix order of arbitrary root pair;
3.   check for odd-labeled cycles via DFS;
4.   swap pairs with odd-labeled root-path;

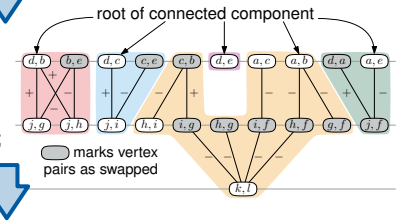

root of connected component

marks vertex pairs as swapped

**Algorithm 2 [2]**
1. **for** each pair, ordered first by level and then by asc. distance of vertices in $\mathcal{L}$ **do**
2.   **if** first-encountered pair for con. comp. **then**
3.     fix according to $\mathcal{L}$;
4.   **else if** neither edges to later pairs nor match with remainder of comp. **then**
5.     exchange in $\mathcal{L}$;
6.   **else if** edge to later pair with label '−' **then**
7.     exchange in $\mathcal{L}$;

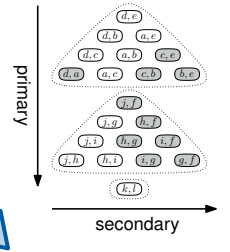

primary / secondary

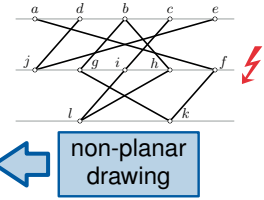

⚡ non-planar drawing

---

**Summary:** We have counterexamples to
- the embedder by Randerath [6]
- the correctness proof of Theorem A [6] (relies on the embedder)
- the embedder by Harrigan & Healy [2] / Healy & Kuusik [3]
- the correctness proof of Theorem B [3] (assumes that eq. classes can always be combined without contradiction)

To the best of our knowledge, this leaves
- no correct *simple* embedding algorithm
- no correct embedder implementation

Disclaimer: Theorem A and B can still be used for simple and correct quadratic tests for Level Planarity, as [1] shows their equivalence to the Hanani-Tutte Level Planarity characterization.

more infos & our interactive implementation of [2]



\end{document}